%% file: main.tex
\documentclass[fleqn,usenatbib]{mnras}

\usepackage{newtxtext,newtxmath}

\usepackage[T1]{fontenc}

\DeclareRobustCommand{\VAN}[3]{#2}
\let\VANthebibliography\thebibliography
\def\thebibliography{\DeclareRobustCommand{\VAN}[3]{##3}\VANthebibliography}

\usepackage{graphicx}	
\usepackage{amsmath}	

\def\arcsec{\hbox{$^{\prime\prime}$}}
\newcommand{\fhs}{\mbox{\ensuremath{.\!\!^{\rm s}}}}

\newcommand{\frbseven}{FRB~20210410D}

\newcommand{\halflight}{\ensuremath{\phi}}
\newcommand{\PU}{\ensuremath{P(U)}}
\newcommand{\PO}{\ensuremath{P(O)}}
\newcommand{\POx}{\ensuremath{P(O|x)}}

\newcommand{\hostname}{J214420.69$-$791904.8}

\newcommand{\dmunits}{\ensuremath{{\rm pc \, cm^{-3}}}}
\newcommand{\dmcosmic}{\ensuremath{{\rm DM}_{\rm cosmic}}}
\newcommand{\dmacosmic}{\ensuremath{\langle {\rm DM}_{\rm cosmic} \rangle}}

\newcommand{\dmhost}{\ensuremath{{\rm DM}_{\rm host}}}

\newcommand{\dmhalo}{\ensuremath{{\rm DM}_{\rm halo}}}
\newcommand{\dmism}{\ensuremath{{\rm DM}_{\rm ISM}}}

\newcommand{\fdz}{\ensuremath{f_d(z)}} 

\title[A sub-arcsec localised FRB with a significant host DM]{A sub-arcsec localised fast radio burst with a significant host galaxy dispersion measure contribution}

\author[M. Caleb et al.]{M. Caleb,$^{1, 2, 3}$\thanks{E-mail: manisha.caleb@sydney.edu.au}
L. N. Driessen,$^{4, 1}$
A. C. Gordon,$^{5}$
N. Tejos,$^{6}$
L. Bernales,$^{6}$
H. Qiu,$^{7}$ 
J.~O. Chibueze,$^{8, 9, 10}$
\newauthor
B. W. Stappers,$^{2}$
K. M. Rajwade,$^{11,2}$
F. Cavallaro,$^{12,13}$
Y. Wang,$^{1, 14, 15}$
P. Kumar,$^{16, 17}$
W. A. Majid,$^{18,19}$
\newauthor
R. S. Wharton,$^{18}$
C. J. Naudet,$^{18}$
M. C. Bezuidenhout,$^{8}$
F. Jankowski,$^{2,20}$
M. Malenta,$^{2}$
V. Morello,$^{2}$
\newauthor
S. Sanidas,$^{2}$
M. P. Surnis,$^{2, 21}$
E. D. Barr,$^{22}$
W. Chen,$^{22}$
M. Kramer,$^{22,2}$
W. Fong,$^{5}$
C. D. Kilpatrick,$^{5}$
\newauthor
J. Xavier Prochaska,$^{23, 24}$
S. Simha,$^{23}$
C. Venter,$^{8}$
I. Heywood,$^{25, 26}$
A. Kundu,$^{8}$
F. Schussler$^{27}$
\\ \\
$^{1}$ Sydney Institute for Astronomy, School of Physics, The University of Sydney, NSW 2006, Australia\\
$^{2}$ Jodrell Bank Centre for Astrophysics, University of Manchester, Oxford Road, Manchester M13 9PL, UK\\
$^{3}$ ASTRO3D: ARC Centre of Excellence for All-sky Astrophysics in 3D, ACT 2601, Australia\\
$^{4}$ CSIRO, Space and Astronomy, PO Box 1130, Bentley, WA 6102, Australia\\
$^{5}$ Center for Interdisciplinary Exploration and Research in Astrophysics (CIERA) and Department of Physics and Astronomy, Northwestern University, Evanston, IL 60208, USA \\
$^{6}$ Instituto de F\'isica, Pontificia Universidad Cat\'olica de Valpara\'iso, Casilla 4059, Valpara\'iso, Chile\\
$^{7}$ SKA Observatory, Jodrell Bank, Lower Withington, Macclesfield, SK119FT, UK\\
$^{8}$ Centre for Space Research, North-West University, Potchefstroom Campus, Private Bag X6001, Potchefstroom, South Africa, 2520\\
$^{9}$ Department of Mathematical Sciences, University of South Africa, Cnr Christian de Wet Rd and Pioneer Avenue, Florida Park, 1709, Roodepoort, South Africa\\
$^{10}$ Department of Physics and Astronomy, Faculty of Physical Sciences, University of Nigeria, Carver Building, 1 University Road, Nsukka 410001, Nigeria\\
$^{11}$ ASTRON, the Netherlands Institute for Radio Astronomy, Oude Hoogeveensedijk 4, 7991 PD Dwingeloo, The Netherlands\\
$^{12}$ INAF-Osservatorio Astrofisico di Catania, Via S. Sofia 78, I-95123 Catania, Italy\\
$^{13}$ Inter-University Institute for Data Intensive Astronomy \& Department of Astronomy, University of Cape Town, Private Bag X3,
Rondebosch, 7701, South Africa\\
$^{14}$ CSIRO, Space and Astronomy, PO Box 76, Epping, NSW 1710, Australia \\
$^{15}$ ARC Centre of Excellence for Gravitational Wave Discovery (OzGrav), Hawthorn, Victoria, Australia\\
$^{16}$ Centre for Astrophysics and Supercomputing, Swinburne University of Technology, PO Box 218, Hawthorn, VIC 3122, Australia \\
$^{17}$ Department of Particle Physics \& Astrophysics, Weizmann Institute of
Science, Rehovot 7610001, Israel \\
$^{18}$Jet Propulsion Laboratory, California Institute of Technology, Pasadena, CA 91109, USA\\
$^{19}$Division of Physics, Mathematics, and Astronomy, California Institute of Technology, Pasadena, CA 91125, USA\\
$^{20}$ LPC2E, Universit\'{e} d'Orl\'{e}ans, CNRS, 3A Avenue de la Recherche Scientifique, 45071 Orl\'{e}ans, France\\
$^{21}$ Department of Physics, IISER Bhopal, Bhauri Bypass Road, Bhopal, 462066, India \\
$^{22}$ Max-Planck-Institut f\"ur Radioastronomie, Auf dem H\"ugel 69, D-53121 Bonn, Germany \\
$^{23}$ University of California, Santa Cruz, 1156 High St., Santa Cruz, CA 95064, USA\\
$^{24}$ Kavli Institute for the Physics and Mathematics of the Universe,
5-1-5 Kashiwanoha, Kashiwa, 277-8583, Japan \\
$^{25}$ Department of Physics, Astrophysics, University of Oxford, Denys Wilkinson Building, Keble Road, Oxford OX1 3RH, UK\\
$^{26}$ Department of Physics and Electronics, Rhodes University, PO Box 94, Grahamstown 6140, South Africa\\
$^{27}$ IRFU, CEA, Université Paris-Saclay, F-91191 Gif-sur-Yvette, France\\
}

\date{Accepted XXX. Received YYY; in original form ZZZ}

\pubyear{2022}

\begin{document}
\label{firstpage}
\pagerange{\pageref{firstpage}--\pageref{lastpage}}
\maketitle

\begin{abstract}
We present the discovery of \frbseven\, with the MeerKAT radio interferometer in South Africa, as part of the MeerTRAP commensal project. \frbseven\, has a dispersion measure DM = $578.78\pm2$ \dmunits\, and was localised to sub-arcsec precision in the 2\,s images made from the correlation data products. The localisation enabled the association of the FRB with an optical galaxy at $z = 0.1415$, which when combined with the DM places it above the 3$\sigma$ scatter of the Macquart relation. We attribute the excess DM to the host galaxy after accounting for contributions from the Milky Way's interstellar medium and halo, and the combined effects of the intergalactic medium and intervening galaxies. This is the first FRB that is not associated with a dwarf galaxy, to exhibit a likely large host galaxy DM contribution. We do not detect any continuum radio emission at the FRB position or from the host galaxy down to a 3$\sigma$ RMS of 14.4~$\mu$Jy/beam. The FRB has a scattering delay of $29.4^{+2.8}_{-2.7}$ ms at 1 GHz, and exhibits candidate subpulses in the spectrum, which hint at the possibility of it being a repeating FRB. Although not constraining, we note that this FRB has not been seen to repeat in 7.28~h at 1.3 GHz with MeerKAT, 3~h at 2.4 GHz with Murriyang and 5.7~h at simultaneous 2.3 GHz and 8.4 GHz observations with the Deep Space Network. We encourage further follow-up to establish a possible repeating nature. 
\end{abstract}




\section{Introduction}

Fast Radio Bursts (FRBs) are energetic bursts of radio emission spread over luminosity distances ranging from 131 Mpc \citep{kmn+22} to 6 Gpc \citep{rbb+22}. FRBs typically last a few microseconds to milliseconds in duration with luminosities spanning $10^{38} - 10^{46}$ erg s$^{-1}$ \citep{lml+20}. Their $10^{12}$ times higher luminosities compared to pulsars and rotating radio transients (assuming beamed radiation) suggest extreme neutron star manifestations as progenitors, which could be isolated, or involve interaction or collision. Other progenitor models involving black holes, white dwarfs, and even exotic stars have also been proposed. Synchrotron-maser emission produced at the shock front between a pulsar wind nebula and a supernova remnant or the interstellar medium (ISM), as well as giant flares from within the magnetosphere of the neutron star, are popular progenitor models \citep{pww+19}.
Repeating FRBs have been observed to emit at frequencies from 110~MHz \citep{pmb+21} all the way up to 8~GHz \citep{GSP+18}, whereas apparently non-repeating sources have only been detected from 350 MHz \citep{pck+20} up to 1.4~GHz. However, their emission outside this frequency range remains uncertain despite attempts to detect them beyond these frequencies. 

In the current known sample, 1 in every 14 FRBs have been observed to repeat and whether they all do is very much an open question~\citep[e.g.][]{css18}. 
Both repeating and (as-yet) non-repeating ones are slowly and steadily being localised to (sub)-arcsec precision either through their repeat pulses or interferometric localisations upon detection \citep[e.g.][]{clw+17, bdp+19}. Optical observations of the well-localised ones have resulted in a sample of 24 FRBs with secure host galaxy associations with redshifts in the range $z = 0.03 - 1.02$ \citep{rbb+22, gfk+23}. Presently, the global host galaxy demographics are diverse to say the least. FRBs are seen to arise in starburst to nearly quiescent galaxies and are typically not associated with the nuclei of the hosts \citep{bha+22}. Positional offsets from the centres of the host galaxies range from 0.8~kpc to 20.1~kpc, with the latter being an FRB that resides in a globular cluster in the M81 system \citep{bha+22}. A handful of galaxy images at high spatial resolution show the FRBs to be residing in the spiral arms of galaxies \citep{mfs+21}. A sample of nine FRB host galaxies with redshifts below 0.5 asserts a strong correlation between the rotation measures (RMs) of the host galaxies and the estimated host dispersion measure (DM) contributions \citep{mpp+22}. The magnetic fields ($\approx 0.5\, \mu$G) of the sample in \cite{mpp+22} are weaker than those characteristic of the Solar neighbourhood ($\approx 6\, \mu$G). However, they are relatively consistent with a lower limit on the observed range of $2 - 6\, \mu$G for star-forming, disc galaxies. 

Overall, there is no clear distinction between the hosts of repeating and non-repeating FRBs. However, a couple of prolific FRBs, FRBs 20121102A and 20190520B are seen to reside in dwarf galaxies and are associated with persistent emission on compact spatial scales \citep{mph+17, nal+22}. These two FRBs are observed to exhibit temporal RM variations indicative of magnetic fields varying on short timescales, which may correlate with the presence of persistent emission. Furthermore, both these FRBs 
exhibit significant host galaxy DM contributions \citep{tbc+17, nal+22} which are expected to arise from the vicinities of the progenitors and are attributed to the likely dense persistent radio sources associated with them. \cite{rlm+21R} have found statistical evidence for a population of FRBs discovered with CHIME at $z\sim0.4$ with host galaxy DM contributions of $\sim400\,\dmunits$. This may be plausible by some halo gas models \citep{rlm+21R}, as well as augmentation by intervening foreground galaxies \citep{jpm+22}. Host galaxies belonging to massive galaxy clusters have also been seen to contribute to the observed excess extragalactic DMs of two FRB sources  \citep{crc+23}.


More recently, FRB 20210117A discovered by the ASKAP telescope was found to have host galaxy characteristics similar to those of FRBs 20121102A and 20190520B and an excess DM contribution from the host \citep{bgs+22}. However, the FRB has not yet been observed to repeat in 9.2 hours of radio follow-up. While there is no direct evidence of \textit{any} FRB being a true non-repeater, such as being produced from a cataclysmic event, the analysis of a large sample of repeaters and non-repeaters by \cite{pgk+21} has shown clear distinctions in the pulse morphology between the two classes. \cite{pgk+21} show that an FRB can be probabilistically classified as either a one-time event or a repeater burst, based solely on its burst morphology. Bursts from repeating sources, on average, have larger widths and, on average are spectrally band-limited (i.e. narrower in bandwidth). These features can be related to the presence of downward frequency drifting or the `sad-trombone' effect, where subbursts under the FRB pulse envelope are seen to cascade downward towards lower frequencies at later times \citep{hss+19}. This behaviour is observed in almost all repeating FRBs and is likely a combination of the unknown emission mechanism and line-of-sight propagation effects. Based on the known sample of repeating and non-repeating FRBs, it appears that frequency downward-drifting may be predictive of repetition. In some cases, various sub-pulses of the same FRB are even observed to have slightly different DMs \citep{dds+20}.

In this paper, we present the discovery of \frbseven, which bears some of the hallmarks of repeating FRBs, but has not yet been observed to repeat in follow-up radio observations.  In Section \ref{sec:meertrap} we present the observational configuration of the MeerKAT telescope and the details of the transient detection pipeline used by the MeerTRAP project. Section \ref{sec:discovery} reports the discovery and the measured and inferred properties of the FRB. We present our analyses and results in Section \ref{sec:analysis_results} and discuss them in Section \ref{sec:discussion}. A summary and our conclusions are presented oin Section \ref{sec:conclusion}.

\section{The MeerTRAP Project}
\label{sec:meertrap}

The MeerKAT radio telescope is a 64-dish interferometer operated by the South African Radio Astronomy Observatory (SARAO) in the Karoo region in South Africa. The dishes are spread over 8 km, with 40 of them concentrated in the inner $\sim1$-km core. Each telescope is 13.5-m in diameter and currently operates regularly in the L-band (856 -- 1712 MHz) and the UHF-band (544 -- 1088 MHz). The MeerTRAP project is a complementary programme to search for pulsars and fast transients while piggybacking on the large survey programmes of MeerKAT. MeerKAT simultaneously observes in incoherent and coherent modes using the MeerTRAP backend. The MeerTRAP backend is the association of two systems: the Filterbank and Beamforming User Supplied Equipment (FBFUSE), a many-beam beamformer that was designed and developed at the Max-Planck-Institut f{\"u}r Radioastronomie in Bonn~\citep{barr2018, CBK+21}, and the Transient User Supplied Equipment (TUSE), a real-time transient detection instrument developed by the MeerTRAP team at the University of Manchester.
When operating at L-band in the coherent mode, the voltages from the inner 40 dishes of the $\sim1$-km core of the array are coherently combined to form up to 780 beams on the sky with an aggregate field-of-view (FoV) of $\sim0.4$~deg$^{2}$ (i.e. overlap at 25\% of the peak power). In the incoherent mode, the intensities of all available MeerKAT dishes (up to a maximum of 64) are added to create a less sensitive but much wider FoV of $\sim$1.3~deg$^{2}$ \citep{rsw+21}. We utilise the highly optimised Graphics Processing Unit (GPU)-based \textsc{astroaccelerate} software \citep{akg+11, aa20} to search for dispersed signals. In the L-band the real-time search is performed by incoherently de-dispersing in the DM range 0--5118.4~\dmunits and searching up to maximum boxcar widths of 0.67\,s. The extracted candidate files contain raw filterbank data of the dispersed pulse and additional padding of 0.5\,s at the start and at the end of the file. See \cite{chr+22, rbc+22, 2023Jankowski} for more details. Additionally, correlated visibilities from as many dishes as are available are recorded during each observation to search for and identify serendipitous transients in images of the field.

\section{Discovery}
\label{sec:discovery}

\begin{figure*}
 \centering
\includegraphics[width=7in]{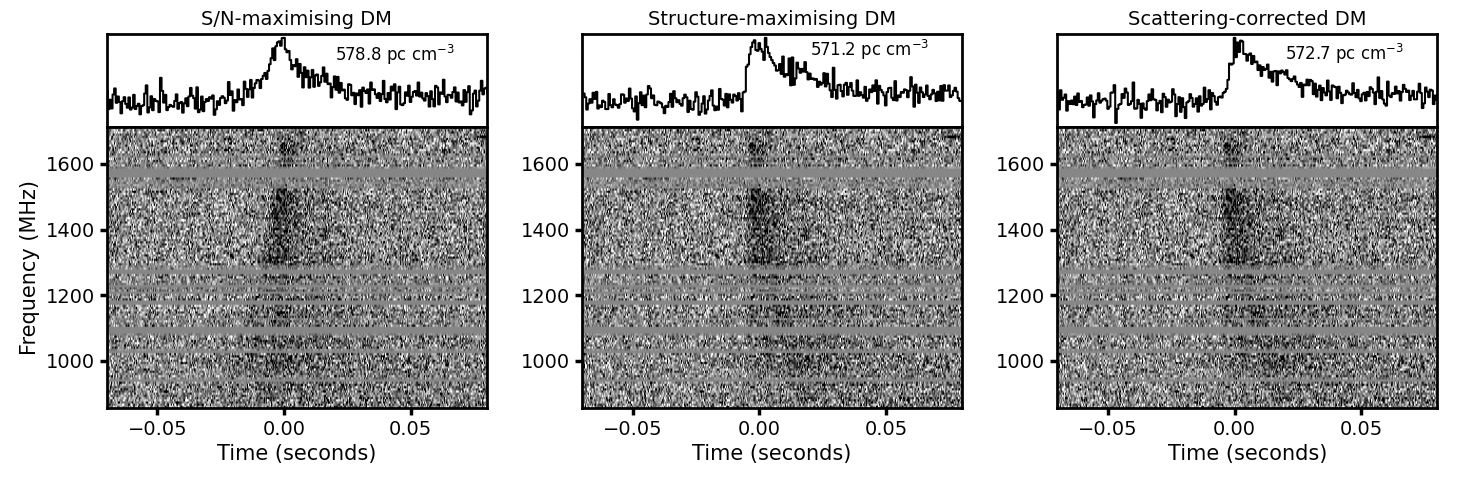}
\caption{Dynamic spectra of the FRB de-dispersed to the S/N-maximising DM (left panel), structure-maximising DM (middle panel) and scattering-corrected DM. The top panel of each case shows the frequency-averaged pulse profile. The data are uncalibrated and the flux densities are in arbitrary units.}
\label{fig:dynspec} 
\end{figure*}

\frbseven\, shown in Figure \ref{fig:dynspec} was discovered while commensally observing with a MeerKAT open-time proposal (SCI-20210212-CV-01; PI Venter) to identify persistent radio emission associated with well-localised FRBs \citep{JOC+22}. It was detected by the MeerTRAP real-time transient pipeline in the incoherent beam while observing the position of FRB~20190611B. In offline analysis, we optimized the signal-to-noise (S/N) of \frbseven\, to $\sim 42$ for a dispersion measure (DM) of 578.78\,\dmunits\ and width of 26.6~ms. See Table \ref{tab:burstproperties} for more details. As the FRB was detected in the incoherent beam, it could lie anywhere within $\sim1.3$ deg$^{2}$. However, since it was bright, we were able to localise it to (sub)-arcsec precision in the simultaneous shortest timescale correlated visibilities. See Section \ref{sec:localisation} for details.

\begin{figure}
    \centering
    \includegraphics[width=\columnwidth]{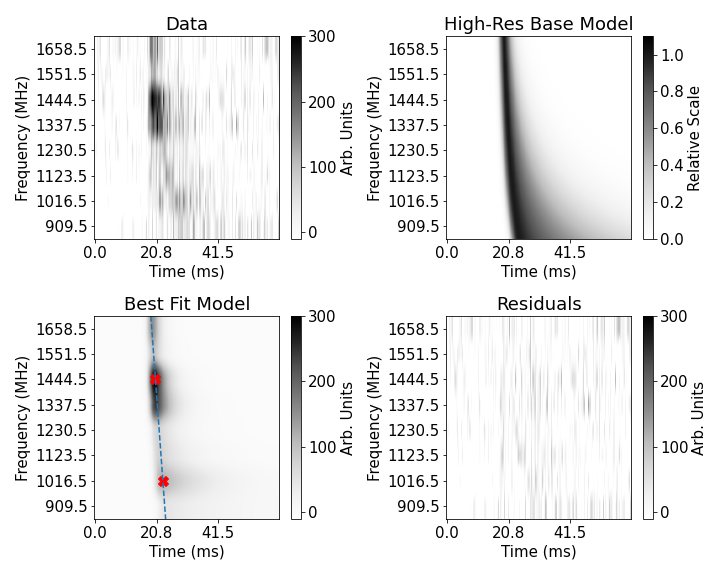}
    \caption{Multi-frequency modelling of the pulse profile dedispersed to the structure-maximising DM. We transform the pulse profile into 8 subbands and apply an 8 subband multi-frequency scattering single pulse model to fit the data. Our results show that a shifted scattering model best describes the pulse profile of \frbseven. The two candidate subpulse positions are marked in the best fit model as crosses with the measured drift rate drawn as the dashed line. We display the data, best-fit model and residuals for comparison.}
    \label{fig:scattering}
\end{figure}

\section{Analysis}
\label{sec:analysis_results}

\subsection{Beamformed radio data analysis}

\subsubsection{DM estimation}

The relatively wide width of \frbseven\, suggests a repeater origin \citep{pgk+21}. Consequently, we calculated the DM to optimise the structure under the burst envelope using \texttt{DM\_phase} (\url{https://github.com/danielemichilli/DM\_phase}). The \texttt{DM\_phase} algorithm finds the structure-optimised DM of a burst by maximising the coherent power across the bandwidth \citep{Seymour_2019}. We dedispersed the data over a trial DM range of $565.0 \leq \rm{DM} \leq 580.0$ \dmunits\, in steps of 0.1 \dmunits. The uncertainty on each DM estimate was calculated by converting the standard deviation of the coherent power spectrum into a standard deviation in DM via the Taylor series. We measure a structure-optimised DM of $571.2 \pm 1$ \dmunits\ for \frbseven. Visual inspection of the spectrum after dedispersing to the structure-optimised DM showed hints of scattering.

\begin{figure}
    \centering
    \includegraphics[width=\columnwidth]{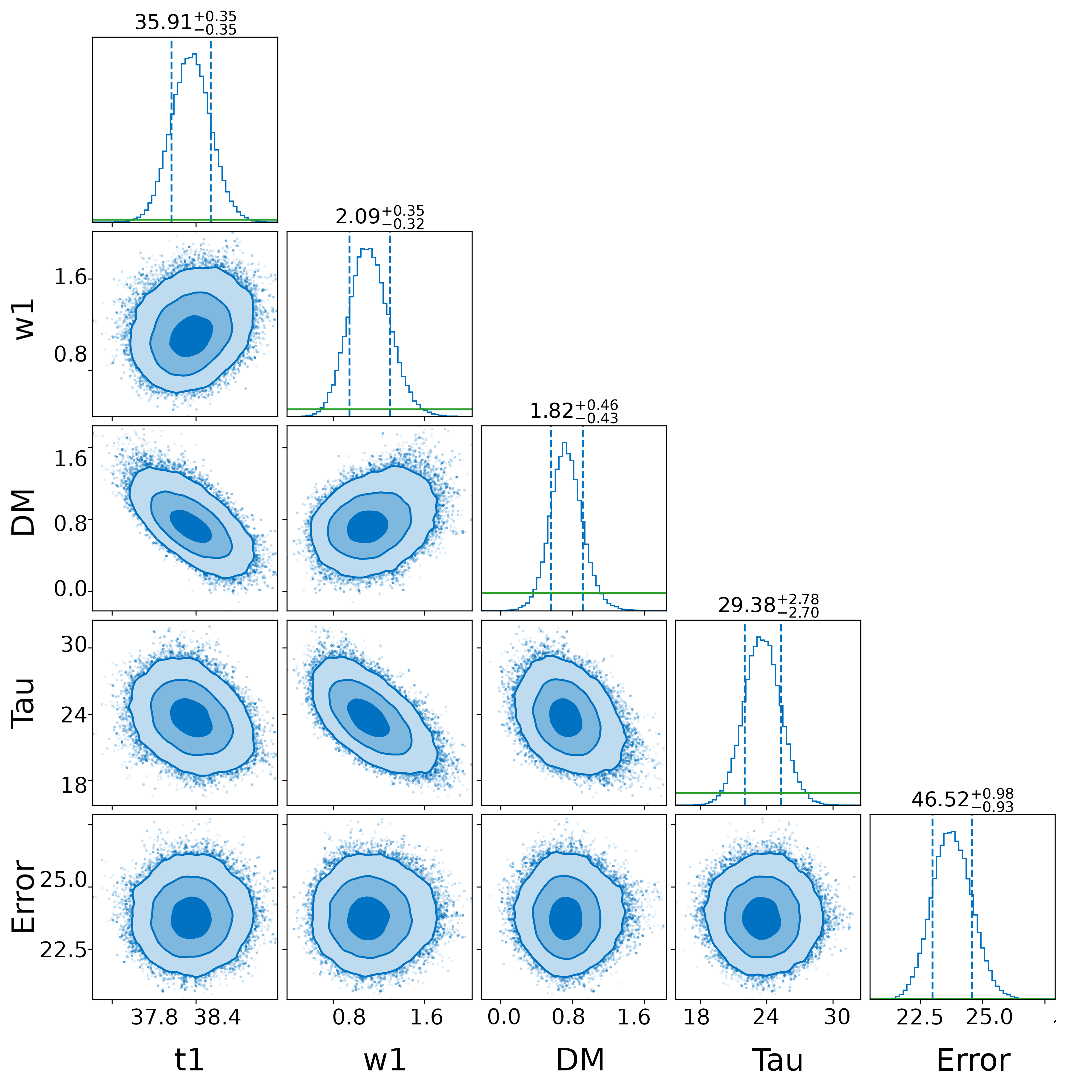}
    \caption{The posterior for modelling the burst dynamic spectrum. We display the key parameters: pulse position (t1), intrinsic width (w1), excess DM (DM) and scattering ($\tau$). The three shades of contour correspond to $1\sigma$, $2\sigma$, and $3\sigma$ confidence levels.}
    \label{fig:posterior}
\end{figure}

\begin{table*}
\caption{Observed and inferred properties of \frbseven\, and the associated host galaxy.}
\label{tab:burstproperties}
\begin{tabular}{lcc}
\hline
Parameter                   &     Unit                              & FRB 20210410\\ 
\hline
\multicolumn{3}{c}{Measured}\\
MJD$_\text{topo}^\text{a}$  &                               & 59314.4673892632 \\
UTC$_\text{topo}^\text{a}$  &                               & 2021-04-10 11:13:02.432\\
Beam                        &                               & Incoherent beam \\
RA                          & (hms)                         & 21$^{\rm h}$44$^{\rm m}$20\fhs7s$\pm$0\farcs8 \\
Dec                         & (dms)                         & $-79^{\mathrm{\circ}}$19\arcmin05\farcs5$\pm$0\farcs5.  \\
$l$                     & (deg)                             & 312.3222099  \\
$b$                     & (deg)                             & $-$34.1293941\\
S/N-maximising DM       & ($\text{pc} \: \text{cm}^{-3}$)   & 578.78 $\pm$ 2.0\\
Structure-maximising DM & ($\text{pc} \: \text{cm}^{-3}$)   & 571.16 $\pm$ 0.97 \\
Scattering-corrected DM & ($\text{pc} \: \text{cm}^{-3}$)   & 572.65 $\pm$ 0.38 \\
$\text{S/N}$            &                                   & 42 \\
Scattering time, $\tau_\mathrm{s}$ at 1~GHz & (\text{ms}) & $29.4^{+2.8}_{-2.7}$\\
\hline
\multicolumn{3}{c}{Inferred}\\
$S_\text{peak}^\text{b}$     & (Jy)                             & 1.5\\
$F^\text{b}$                 & (Jy ms)                          & 35.4\\
DM$_{\text{MW,NE2001}}$      & ($\text{pc} \: \text{cm}^{-3}$)  & 56.2 \\
DM$_{\text{MW,YMW16}}$       & ($\text{pc} \: \text{cm}^{-3}$)  & 42.2\\
DM$_{\text{halo}}$           & ($\text{pc} \: \text{cm}^{-3}$)  & 40\\
DM$_\text{EG}$               & ($\text{pc} \: \text{cm}^{-3}$)  & 489 \\
\hline

\multicolumn{3}{c}{Host galaxy}\\
Host galaxy name &         & \hostname \\
Redshift ($z$)      &                                   & 0.1415 \\
g & (AB mag) & 21.77$\pm$0.05 \\
r & (AB mag) & 20.65$\pm$0.03 \\
i & (AB mag) & 20.10$\pm$0.02 \\
z & (AB mag) & 20.23$\pm$0.04 \\
Y & (AB mag) & 19.76$\pm$0.16 \\
J & (AB mag) & 20.02$\pm$0.21 \\
Stellar metallicity & log (Z$_*$/Z$_{\odot}$) & $-1.08^{+0.21}_{-0.33}$  \\
Stellar mass & log (M$_*$/M$_{\odot}$) & $9.46^{+0.05}_{-0.06}$ \\
0-100 Myr Integrated SFR & (M$_{\odot}$ yr$^{-1}$) & $0.03^{+0.03}_{-0.061}$ \\
Projected offset from galaxy centre & (kpc) & 2.9\\
H$_\alpha$ flux density & (erg s$^{-1}$ cm$^{-2}$) & $1.5\times 10^{-16}$ \\

\hline
\multicolumn{3}{l}{$^\text{a}$ Topocentric arrival times measured at the highest frequency channel, 1711.58~MHz.}\\

\end{tabular}
\end{table*}

\subsubsection{Scattering Analysis}
\label{sec:scattering}
We performed a 2D scattering fit to the FRB using a python-based burst model software and Monte Carlo sampling methodology based on \cite{qiu2020}. We utilise a robust dynamic modelling in a higher resolution model to account for the convolved scatter broadening, dispersion smearing and intrinsic pulse width. The model also takes into account of any possible varying pulse arrival time across frequency due to incomplete dedispersionas seen in e.g \citet{hss+19}. For FRB 20210410D, we assume a single Gaussian pulse model with scatter broadening. To avoid confusion with intrinsic pulse width and also to accurately determine scattering caused by inhomogeneous plasma, we fix the scattering index to $\alpha=-4$.

We measure a scattering time of $29.4^{+2.8}_{-2.7}$ ms at 1 GHz and an intrinsic pulse width of $1\sigma = \mathrm{2.0\pm0.3\,ms}$. The pulse position is delayed in relation to frequency when structure maximised. This can be characterised as an excess dispersion delay of $1.8^{+0.5}_{-0.4}$\,\dmunits, which corresponds to the scattering corrected DM. The pulse position shift can also be characterised by a drift rate of $1.2 \pm 0.4\ \mathrm{ms/107\ MHz}$ by measuring the two subband positions at 1444.5~MHz and 1016.5~MHz. Our relatively coarse time resolution of 306.24 $\upmu$s does not allow us to robustly distinguish between the two. We show our best-fit model and residuals in Figure \ref{fig:scattering} and the parameter posteriors in Figure \ref{fig:posterior}.



\subsection{Imaging radio data analysis}
\subsubsection{\frbseven\, localisation}
\label{sec:localisation}

The shortest integration image we could make during the 2021 April 10 observation when \frbseven\, was detected was 2\,s. Though the standard integration time for MeerKAT imaging data is 8\,s, it is possible to go to shorter
integration times for certain projects. The MeerKAT open-time proposal SCI-20210212-CV-01 used an integration time of 2\,s for their data, as there was a possibility of detecting repeat pulses from the target FRB~20190611B. 
As such, we imaged the observations into 2\,s chunks. The 2\,s images have a typical root-mean-square (RMS) noise of $\sim0.7$ mJy $\mathrm{beam^{-1}}$ and a synthesised beam size of 21\arcsec$\times$10\arcsec.

\begin{figure*}
 \centering
\includegraphics[width=5.0 in]{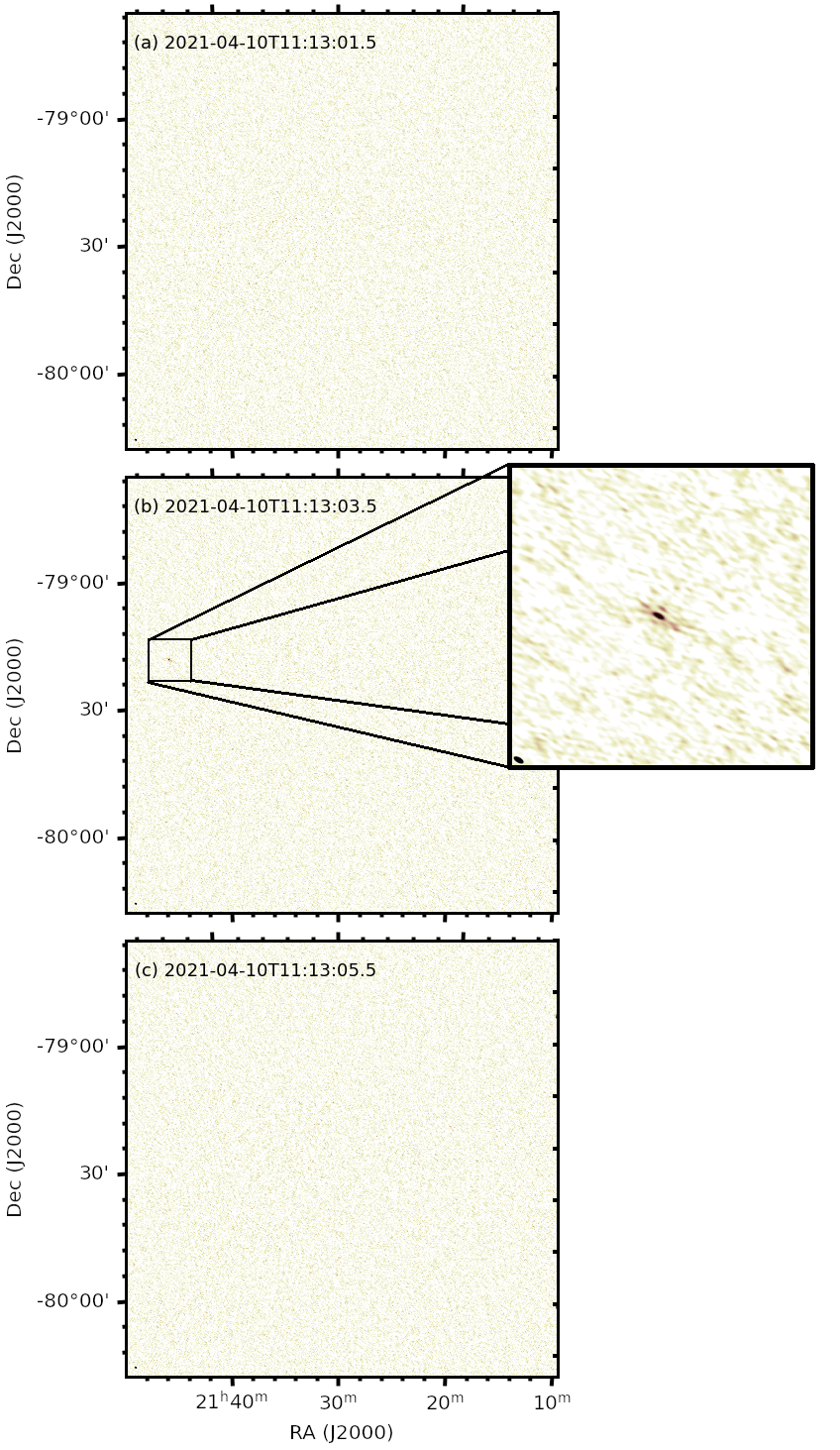}
\caption{Difference images of (a) the time step before, (b) the time step of, and (c) the time step after the FRB detection. The time steps shown are the centre of the time slice. The colour scale is the same in each panel and the synthesised beam is shown in the bottom left corner of each panel and the inset image. The inset image shows a 1\arcmin$\times$1\arcmin\, zoom in of the FRB position. No other variables are detected in this field in these time steps.}
\label{fig: difference images} 
\end{figure*}

To find the position of the FRB we used difference imaging. The total dispersion delay of~\frbseven~across the MeerKAT band is $\sim$2.4\,s and based on the time of detection, we expect most of the FRB to lie within a single 2\,s image. We use {\tt WSClean} to produce $20\times2$\,s images around the time of the MeerTRAP FRB detection. We then combined these images into an average image and subtracted this average image from each individual 2\,s image. The subtracted images of the time step before, the time step of, and the time step after the FRB detection are shown in Figure\,\ref{fig: difference images}. We can see in this figure that only the difference image at the time of the detection has a bright spot, which is $\sim$40\arcmin\, from the phase centre of the images. As this source in the difference image is detected at the time that we expect to see the FRB, it is outside of the MeerTRAP coherent beam tiling as expected, it is the only source in the difference image, and there is no source in the difference images surrounding the expected FRB detection time, we determine that this source is the FRB. The FRB has a flux density of 35 mJy with a $\sim1.8$ mJy RMS.

The position of the FRB detected in the difference image prior to astrometric correction is 21$^{\rm h}$44$^{\rm m}$20\fhs6s$\pm$0\farcs7 $-79^{\mathrm{\circ}}$19\arcmin04\farcs8$\pm$0\farcs3. 

\subsubsection{Absolute astrometry}
\label{sec:astrometry}

We corrected the astrometry in the MeerKAT images using the method described in \citet{2022MNRAS.512.5037D}. We imaged the full-time integration image of 2.25~hours to determine the astrometric correction. We first find the positions of the sources in the images using the Python Blob Detector and Source Finder\footnote{\href{https://www.astron.nl/citt/pybdsf/}{https://www.astron.nl/citt/pybdsf/}} ({\tt pyBDSF}). We then find Australian Telescope Compact Array (ATCA) Parkes-MIT-NRAO (PMN) \citep[ATPMN;][]{2012MNRAS.422.1527M} sources within the MeerKAT FoV, excluding any sources that appear resolved in the MeerKAT images. There are 27 ATPMN sources in the FoV, 13 of which are resolved. The ATPMN positions of the unresolved sources, and the separations between the ATPMN and MeerKAT sources before and after applying the astrometric correction are shown in Table\,\ref{tab: ATPMN sources}. The ATPMN survey has a median absolute astrometric uncertainty of 0\farcs4 in both Right Ascension (RA) and Declination (Dec). We use the matching ATPMN and MeerKAT point sources to solve for a transformation matrix to shift and rotate the MeerKAT sources to match the ATPMN source positions\footnote{The code for performing the astrometric corrections can be found on GitHub: \href{https://github.com/AstroLaura/MeerKAT\_Source\_Matching}{https://github.com/AstroLaura/MeerKAT\_Source\_Matching}}. We apply the transformation matrix to all MeerKAT source positions and we combine the {\tt pyBDSF} uncertainties and 0\farcs4 ATPMN uncertainty in quadrature. In shorter time-scale MeerKAT imaging, 2\,s or 8\,s images, we often do not detect many ATPMN sources. As such, we determine the astrometric correction using the sources extracted from the full integration image of the epoch and apply that correction to the sources extracted from the 2 or 8\,s images.  We find that the corrected position of the FRB is 21$^{\rm h}$44$^{\rm m}$20\fhs7s$\pm$0\farcs8 $-79^{\mathrm{\circ}}$19\arcmin05\farcs5$\pm$0\farcs5.

\begin{table*}
    \centering
    \begin{tabular}{llrrrr}
    & & \multicolumn{2}{c}{2021 Apr 10 Offset} & \multicolumn{2}{c}{2021 Sep 05 Offset} \\
    RA & Dec & Before correction & After correction & Before correction & After correction \\
    \hline
    \hline
20$^{\rm h}$57$^{\rm m}$06\fhs5 & -80$^{\mathrm{\circ}}$18\arcmin24\farcs3 & 1\farcs3 & 0\farcs84 & 0\farcs96 & 0\farcs17 \\
20$^{\rm h}$57$^{\rm m}$06\fhs5 & -80$^{\mathrm{\circ}}$18\arcmin24\farcs3 &  &  & 1\farcs5 & 0\farcs51 \\
21$^{\rm h}$05$^{\rm m}$45\fhs0 & -78$^{\mathrm{\circ}}$25\arcmin34\farcs5 & 1\farcs9 & 0\farcs87 & 1\farcs0 & 0\farcs2 \\
21$^{\rm h}$21$^{\rm m}$40\fhs8 & -78$^{\mathrm{\circ}}$03\arcmin46\farcs7 & 1\farcs3 & 0\farcs49 & 0\farcs94 & 0\farcs44 \\
21$^{\rm h}$24$^{\rm m}$40\fhs9 & -80$^{\mathrm{\circ}}$05\arcmin01\farcs0 & 0\farcs83 & 0\farcs64 & 0\farcs25 & 0\farcs5 \\
21$^{\rm h}$24$^{\rm m}$40\fhs9 & -80$^{\mathrm{\circ}}$05\arcmin01\farcs0 &  &  & 0\farcs96 & 0\farcs41 \\
21$^{\rm h}$46$^{\rm m}$30\fhs0 & -77$^{\mathrm{\circ}}$55\arcmin54\farcs7 & 1\farcs3 & 0\farcs69 & 1\farcs6 & 0\farcs8 \\
21$^{\rm h}$47$^{\rm m}$05\fhs8 & -78$^{\mathrm{\circ}}$12\arcmin21\farcs9 & 0\farcs73 & 0\farcs24 & 0\farcs63 & 0\farcs15 \\
21$^{\rm h}$49$^{\rm m}$30\fhs7 & -80$^{\mathrm{\circ}}$46\arcmin02\farcs5 & 1\farcs5 & 1\farcs0 & 0\farcs4 & 0\farcs11 \\
21$^{\rm h}$52$^{\rm m}$03\fhs2 & -78$^{\mathrm{\circ}}$07\arcmin06\farcs4 & 0\farcs58 & 0\farcs21 & 0\farcs49 & 0\farcs39 \\
21$^{\rm h}$56$^{\rm m}$47\fhs8 & -79$^{\mathrm{\circ}}$37\arcmin39\farcs1 & 0\farcs97 & 0\farcs58 & 0\farcs35 & 0\farcs2 \\
21$^{\rm h}$59$^{\rm m}$01\fhs2 & -80$^{\mathrm{\circ}}$39\arcmin16\farcs8 & 1\farcs8 & 1\farcs6 & 0\farcs46 & 0\farcs13 \\
21$^{\rm h}$59$^{\rm m}$16\fhs6 & -80$^{\mathrm{\circ}}$41\arcmin43\farcs8 & 1\farcs5 & 0\farcs88 & 0\farcs6 & 0\farcs54 \\
22$^{\rm h}$06$^{\rm m}$11\fhs2 & -79$^{\mathrm{\circ}}$35\arcmin11\farcs8 & 1\farcs7 & 1\farcs2 & 0\farcs4 & 0\farcs19 \\
    \hline
    \end{tabular}
    \caption{Coordinates of the ATPMN sources used as reference sources for the astrometric corrections. Only unresolved sources were used for the astrometry. Some ATPMN sources were not within the 2021 April 10 observation FoV but were within the 2021 September 05 observation FoV. The offsets are the separations between the ATPMN and MeerKAT positions before and after applying the astrometric corrections.}
    \label{tab: ATPMN sources}
\end{table*}

\subsubsection{Continuum source localisation}
\label{sec:MKTPRS}

The FRB was detected in the $\sim 1$-hour long observation on 2021 April 10, $\sim$40\arcmin\,from the phase centre. This means that the sensitivity to faint, persistent emission near the FRB was low. A 3-hour long observation on 2021 September 05 was pointed to accommodate three FRBs within the MeerKAT FoV. This meant that the position of \frbseven\, was $\sim$28\arcmin\,from the phase centre resulting in higher sensitivity at the FRB position than in the detection observation. On 2021 September 05 a faint, $\sim8\sigma$ (RMS = $\sim4.8 \,\mu$Jy) persistent continuum source was detected $\sim$3\arcsec\,from the FRB position. We corrected the astrometry of the image (the offsets between the MeerKAT positions and ATPMN positions before and after correction are shown in Table\,\ref{tab: ATPMN sources}) and found the corrected position of the persistent continuum source to be 21$^{\rm h}$44$^{\rm m}$19\fhs5s$\pm$0\farcs6 $-79^{\mathrm{\circ}}$19\arcmin05\farcs8s$\pm$0\farcs5 (uncorrected: 21$^{\rm h}$44$^{\rm m}$19\fhs4s$\pm$0\farcs4 $-79^{\mathrm{\circ}}$19\arcmin05\farcs4s$\pm$0\farcs3). The separation between the corrected FRB position and the corrected persistent continuum source position is 3\farcs3. The MeerKAT persistent continuum source is consistent with the position of the galaxy at $z\sim0.4$ (see Section \ref{sec:optical-analyses} for details) as shown in Figure \ref{fig:optical_galaxies} and is not a compact persistent radio source (PRS) as observed in a couple of repeating FRBs.

\subsubsection{Polarization analysis}

\begin{figure}
\centering
\includegraphics[width=3.3 in]{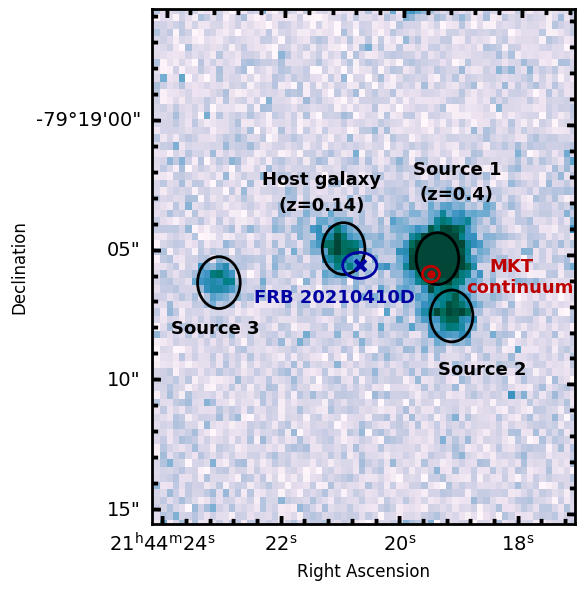}
\caption{200~s $r$-band exposure using the Goodman High Throughput Spectrograph (GHTS) on the SOAR 4.1m telescope. The blue cross and ellipse represent the position of the FRB and the 2$\sigma$ total uncertainty in the position. The red circle and ellipse represent the position of the continuum detection in the deep MeerKAT image and the 2$\sigma$ total uncertainty in the position (see Section \ref{sec:MKTPRS}). Black circles are 1\arcsec\ in diameter and highlight other sources in the field. The absence of emission lines for sources 2 and 3 prevented an estimate of their redshifts.}
\label{fig:optical_galaxies}
\end{figure}

The FRB discovery was made while commensally observing with a MeerKAT open-time proposal that did not require polarization information. Consequently, a standard polarization calibrator was not observed. To extract the polarization information we used, in the 2\,s image containing the FRB detection, J1619$-$8418 as the secondary and polarization calibrator. It is a known calibrator with very low linear ($\sim0.4\%$) and circular ($\sim0.03\%$) polarization\footnote{https://archive-gw-1.kat.ac.za/public/meerkat/MeerKAT-L-band-Polarimetric-Calibration.pdf}. We mapped Stokes Q, U, and V along with Stokes I using the IDIA pipeline\footnote{https://idia-pipelines.github.io/}. \frbseven\, was only detected in Stokes I and not in Stokes Q, U or V, giving us a $3\sigma$ upper limit of about $18\%$ for each of them. We used the \texttt{RMsynth3D} algorithm from the \texttt{RM-tools}\footnote{https://github.com/CIRADA-Tools/RM-Tools} package to perform 3D RM synthesis on the image-frequency cube. The method transforms polarized intensity as a function of $\lambda^{2}$ to Faraday depth, $\phi$, representing polarized intensity for different trial RMs in the range [$-150000, +150000$]. We did not detect a peak in the maximum polarized intensity map at the position of the FRB. We do, however, detect a peak at a $6\sigma$ significance with an RM of $-77,929$ rad m$^{-2}$.
This peak is not associated with any of the sources detected in the field in Figure \ref{fig:optical_galaxies}. 
Therefore, we consider this a spurious detection.

\subsection{Radio follow-up for repeat bursts}

\subsubsection{MeerKAT}
FRB\,20210410D is approximately $1\degr$ away from the position of FRB\,20190611B and $0.7\degr$ away from FRB\,20190102C, both of which were observed as part of the open-time proposal with MeerKAT. The MeerTRAP backend was used to piggy-back these imaging observations to look for any repeat bursts from FRB\,20190611B, FRB\,20190102C and consequently, from FRB\,20210410D in real-time. No other radio bursts (whether repeat bursts from the known FRBs in the field or completely unassociated) were detected in our beamformed data down to a fluence limit of $0.09$ Jy ms in a total of 7.28\,h of time spent on the FRB source with MeerKAT. 

\subsubsection{Murriyang}
In addition, we performed follow-up observations of the FRB\,20210410D source using the ultra-wideband low (UWL) receiver at the 64-m Murriyang (formerly known as Parkes) radio telescope. The observations spanned a bandwidth of 0.7--4\,GHz and were centred at 2368 MHz. We observed the FRB source for a total of 1.9\,h on 2022 January 29 and 1.1\,h on 2022 July 19. We searched the Parkes data for repeat bursts using the GPU-based single-pulse search software \texttt{Heimdall} \citep{Barsdell:2012PhDT} and utilized the neural-network trained model \texttt{Fetch} \citep{Agarwal:2020} for the classification of candidates. We employed a tiered sub-band strategy as described in \citet{Kumar:2021} to search the wide-band Parkes data in the DM range of 100--1100\,pc\,cm$^{-3}$. In these searches, we did not find any significant repetition candidate above an S/N of 8. Using the radiometer equation and assuming a nominal burst width of 1\,ms, we can constrain the detectable fluence of the repeat bursts. Our search pipeline was sensitive up to $\sim$1 Jy\,ms, assuming the repeat bursts are narrowband $\sim64$\,MHz and  $\lesssim$ 0.15 Jy\,ms for a burst spanning the entire UWL band. 

\subsubsection{Deep Space Network}

We also observed FRB 20210410D for a total of $\sim$5.7 hrs over four separate epochs in 2022 (September 3, October 6, October 15, and October 22) with DSS-43, a 70-m diameter dish located at the Deep Space Network’s (DSN) complex in Canberra, Australia. These observations were carried out simultaneously at S-band (centre frequency of 2.3 GHz) and at X-band (centre frequency of 8.4 GHz). The data were recorded in both left and right circular polarization modes using the resident pulsar backend in filterbank search mode, where channelized power spectral density measurements are recorded with 1 MHz channel spacing and a time resolution of 512 $\mu$s. The S-band system spans roughly 120 MHz of bandwidth, while the X-band system spans a bandwidth of 400 MHz.  

The data processing procedures followed the same steps outlined in previous DSN studies of pulsars and FRBs \citep[e.g.][]{2021Majid}. Datasets in each observing band were first corrected for bandpass slope.  We excised bad frequency channels corrupted by radio frequency interference (RFI) using PRESTO’s rfifind package.  To remove any long-timescale temporal variability, we subtracted a 5-s moving average from each data point. Datasets from the two orthogonal polarizations were then summed in quadrature.

The cleaned data were then dedispersed with the nominal DM value of 578.78\,\dmunits.  The resulting time series were searched for FRBs using a matched filtering algorithm by convolving individual time series with logarithmically-spaced boxcar functions of widths ranging between 1-300 times the intrinsic time resolution of the pulsar backend. We did not detect any candidate bursts above a S/N threshold of 7 in either frequency band, corresponding to a flux threshold of 0.33 Jy at S-band and 0.19 Jy at X-band for a 1 ms duration burst.

\begin{figure*}
 \centering
\includegraphics[width=3.3 in]{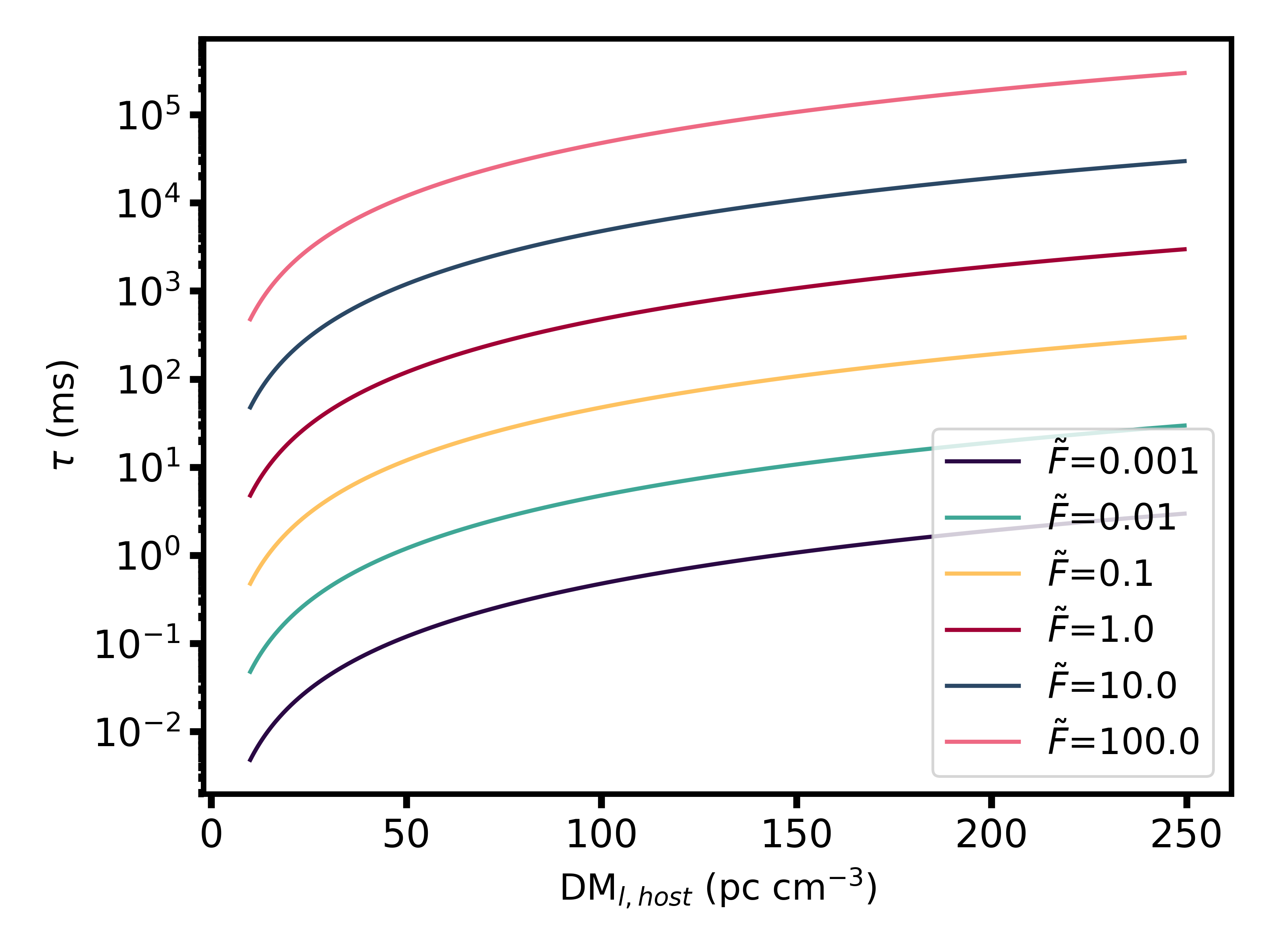}
\includegraphics[width=3.3 in]{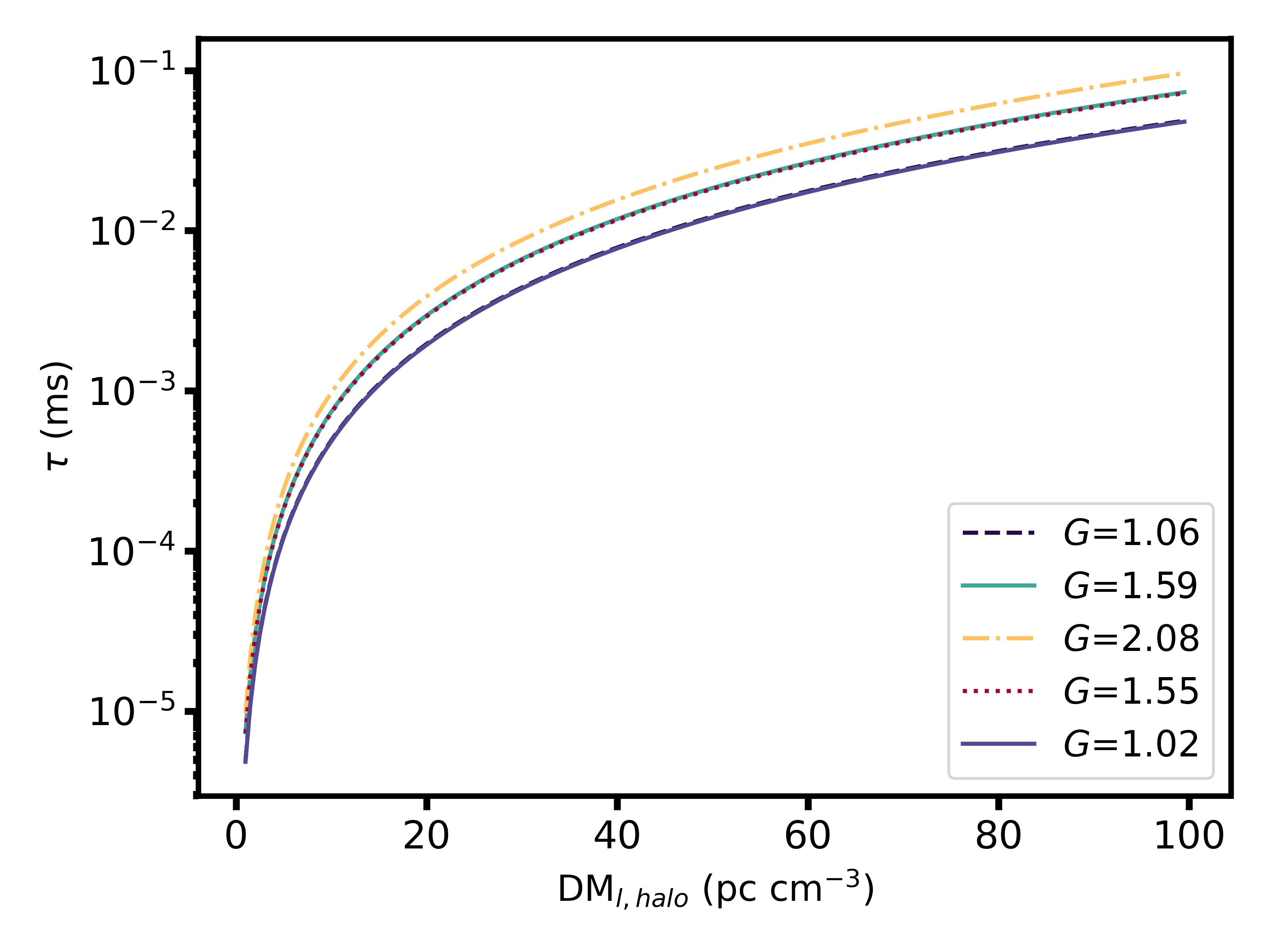}

\caption{\textit{Left Panel}: Scattering time ($\tau$) at 1~GHz as a function of host DM contribution for various values of \protect{$\tilde{F}$} that signifies turbulence for \frbseven. \textit{Right Panel}: \protect{$\tau$} at 1~GHz as a function of DM from the halo of an intervening galaxy for values of $G_\mathrm{scatt}$ (computed from a range of redshifts between 0 and 0.1415) that signifies the geometric boosting to scattering. The analysis is done for a lens with a size of 30~kpc.}
\label{fig:hosthaloscatter} 
\end{figure*}

\subsection{Constraining the location of the scattering screen}

Given the results of the scattering analysis in Section \ref{sec:scattering} we attempt to identify the source of the scattering towards \frbseven. We estimate the scattering timescale from the MW to be 0.28 $\mu$s at 1 GHz using the NE2001 model. The observed scattering time of $29.4^{+2.8}_{-2.7}$ ms at 1 GHz for \frbseven\, is much too large to arise from the MW. Therefore, using the analytical framework provided in \cite{OCC2021}, we estimate the expected scattering from~\frbseven\ in two cases: 1) the scattering originates in the host galaxy of~\frbseven~and 2) the scattering originates in an intervening halo of a foreground galaxy. 
The scattering due to a foreground
galaxy depends not only on its DM contribution but also on the geometric leverage effect, which
will increase the scattering by several orders of magnitude relative to scattering in the host galaxy.
Figure~\ref{fig:hosthaloscatter} shows the expected scattering for both cases as a function of the host galaxy DM contribution and an intervening halo DM contribution. Based on \cite{OCC2021}, the scattering in the host galaxy can be characterised as,

\begin{equation}
\tau (\nu, {\rm DM}, z) \approx 48.03\, \times \, \frac{A_{\tau}\tilde{F}G_\mathrm{scatt}~{\mathrm{DM}_{l}^{2}}}{(1+z_{l})^{3}~\nu^{4}}\, \mu s,    
\end{equation}

\noindent where $A_{\tau}$ is a dimensionless factor approximated to unity, $\tilde{F}$ is the factor that quantifies turbulent density fluctuations in the scattering medium, $\mathrm{DM}_{l}$ is the DM contribution from the scattering screen in either the host galaxy or the intervening halo, $z_{l}$ is the redshift of the scattering screen and $\nu$ is the observing frequency. $G_\mathrm{scatt}$ is a factor that quantifies the geometric enhancement of the scattering due to the distance between the source, the scattering screen, and the observer. Following \cite{OCC2021}, we define it as $G_\mathrm{scatt}\sim\, d_\mathrm{sl}d_\mathrm{lo}/Ld_\mathrm{so}$, where $L$ is the thickness of the thin scattering screen, and $d_\mathrm{sl}$, $d_\mathrm{lo}$ and $d_\mathrm{so}$ are the angular diameter distances of source to lens, lens to observer, and source to observer, respectively. For scattering within the MW or in a distant FRB host galaxy $G_\mathrm{scatt} = 1$, but $G_\mathrm{scatt} \gg 1$ for an intervening halo or galaxy. 

Similar to \cite{rbc+22}, we estimate the scattering time for a foreground galaxy at distances ranging from 25$\%$ to 75$\%$ of the redshift of the host (i.e.) $0.035375 \leq z \leq 0.106125$. For scattering dominated by a screen in the host galaxy, we assume $G\sim$1 and compute the scattering time for various values of $\tilde{F}$. In the case of scattering dominated by the halo or ISM of an intervening galaxy, we conservatively adopt $\tilde{F}\sim 0.0001$ measured for pulsars in the MW~\citep{OCC2021} and calculate the scattering times for various values of $G$ derived from the assumed distances to the screen. For this scenario, we also allow $\tilde{F}$ to evolve with redshift according
to the cosmic star formation history given by using Equation ~21 in~\cite{OCC+22}. According to \cite{OCC+22}, this is based on the assumption that $\tilde{F}$ can evolve with redshift if the underlying turbulence is driven
by star formation feedback or gravitational instability.

\cite{ckr+22} studied the dispersion and scattering properties of a sample of CHIME FRBs and cannot rule out a model of FRBs for which scattering originates in both the local environment and in intervening galaxies.
It is apparent from Figure \ref{fig:hosthaloscatter} that the halos and ISM of intervening galaxies cannot easily account for the scattering time seen in \frbseven\, (see Table \ref{tab:burstproperties}) even for all possible values of $G_\mathrm{scatt}$ within the co-moving distance of the host galaxy of \frbseven. However, the expected $\tau$ at 1~GHz can arise from the host galaxy. We thereby conclude that the scattering seen in this FRB originates from the host galaxy.

\subsection{Optical follow-up}
\label{sec:optical-analyses}

To identify a host galaxy, we observed the FRB position with the SOAR/Goodman High Throughput Spectrograph in $r$-band on 2021 July 19 UT (program ID SOAR2021A-010; PI Fong) and obtained $8\times 200$\,s exposures. Four sources were found surrounding the FRB position, as shown in Figure \ref{fig:optical_galaxies}. Further imaging in the $g$-, $i$-, and $z$-band was performed over several nights spanning 2022 August 10 -- September 03 UT (program ID SOAR2022B-007; PI Gordon), also using $8\times 200$\,s exposures in each band. The data were reduced using the \texttt{photpipe} pipeline \citep{Rest+05}. Each image frame was corrected for bias and flat-fielding following standard procedures using calibration frames obtained on the same night and instrumental configuration.  We then registered the calibrated frames using Gaia DR3 astrometric standards \citep{Gaia} observed in the same field as \frbseven. We performed point-spread function (PSF) photometry on point sources in each image using a custom version of {\tt DoPhot} \citep{Schechter93} and calibrated the photometry with SkyMapper DR2 $griz$ standard stars \citep{skymapper}.  Finally, the individual frames for each band were sky-subtracted, stacked, and regridded to a common field centre and pixel scale with {\tt SWarp} \citep{swarp} using an optimal median weighting derived from the zero point in each frame.

To further fill out the spectral energy distribution, we obtained archival imaging of the field of \frbseven\ from the VISTA Hemisphere Survey (VHS; \citealt{VISTA}) in $Y$- and $J$-band. We used a custom implementation of \texttt{SWarp} \citep{swarp} to stack the reduced data, normalizing them to a zeropoint of 27.5 AB magnitude. We then performed photometry using the same method as described above.

\subsubsection{PATH analysis}

\input{tab_path.tex}

We performed a Probabilistic Association of 
Transients to their Hosts 
\citep[PATH;][]{path} analysis to estimate
posterior probabilities for the host galaxy
candidates of \frbseven.
Our analysis adopts the revised priors
for FRBs (Shannon et al. in prep), 
specifically with an 
exponential scale length of 
1/2 the effective radius.
Furthermore,
we focused the analysis on the 
SOAR/Goodman $r$-band image.
Assuming an unseen prior of 
$\PU = 0$, i.e.\ the host is detected in
our relatively deep image,
we calculate the posterior probabilities
listed in Table~\ref{tab:path}.
It is evident that the galaxy \hostname\ 
which lies closest to the FRB has a
very high posterior probability ($\PO >99.5\%$).
We, therefore, assign it as the host with
high confidence. The offset of \frbseven's position from the galaxy core is not uncommon, as in most cases the FRB localisations are significantly offset from the host galaxy centres \citep{bha+22}.


\subsubsection{Spectroscopic redshift}

We performed longslit spectroscopy with the Gemini-South GMOS spectrograph \citep{gmos, gmos_16} towards \frbseven\ as part of programs GS-2021B-Q-138 and GS-2022A-Q-143 (PI Tejos). Given the relative positions of the sources of interest, we used the 1\arcsec\ slit with two position angles (PAs). On UT 2021 October 14 we used a PA=102\,$\deg$ covering the FRB~host, Source~3 and the (northwestern) outskirts of Source~1, and obtained $3\times 1000$\,s exposures using the R400 grating centred at $700$\,nm. On UT 2022 June 07 we used a PA=10\,$\deg$ covering the centers of Source~1 and Source~2 and obtained $4\times 600$\,s exposures using the R400 grating centred at $700$ and $750$\,nm (two exposures each).\footnote{The use of two central wavelengths was needed in order to cover a CCD gap produced by a reported failure in one of the GMOS-South CCD amplifiers.} A $2\times2$ binning was used in all the exposures.

We reduced these data with the \texttt{PypeIt} pipeline \citep{pypeit} using standard recipes. For the host galaxy, we established a secure redshift of $z=0.1415$ from several emission lines present in the spectrum (H$\alpha$, H$\beta$, [N~{\sc ii}], S~{\sc ii}). For Source~1 we could only identify a single emission line and thus the redshift is not secure; given its observed wavelength we deem this line to be H$\alpha$ at $z=0.4$ (background). For Sources~2~and~3 we could not identify any strong emission line. We note that Source~3 appears spatially unresolved and thus it may well be a star.

These results have been recently confirmed by subsequent VLT/MUSE integral field unit (IFU) observations obtained on UT 2022 October 25 (e.g. Bernales et al. in prep.), which securely confirm the H$\alpha$ identification of Source~1 (and thus its redshift); furthermore, these data have securely established that Source~2 is also background to \hostname.

\subsubsection{Spectral energy distribution analysis}
To understand the stellar population properties of the host galaxy \hostname, we used the Bayesian inference code \texttt{Prospector} \citep{Johnson+21} with a \texttt{continuity} non-parametric star formation history \citep[SFH, ][]{Leja2019}. \texttt{Prospector} performs stellar population synthesis by jointly fitting photometric and spectroscopic data using the \texttt{python-fsps} stellar population synthesis library \citep{Conroy2009, Conroy2010}. The models were then sampled using the dynamic nested sampling routine \texttt{dynesty} \citep{Speagle2020}. Additionally, we implement the \citet{Kroupa01} IMF, \citet{KriekandConroy13} dust attenuation curve, the \citet{Gallazzi2005} mass-metallicity relationship, and a ratio on dust attenuation from young and old stars. To model the spectroscopy, we use a 12th-order Chebyshev polynomial to fit the observed spectrum to the model spectrum. Further, spectroscopic priors include a spectral smoothing parameter, a pixel outlier model to marginalize over noise, and a noise inflation model to ensure a good fit between the observed and model spectrum. Several selected stellar population parameters are presented in Table \ref{tab:burstproperties}.


\subsection{Disentangling the DM contributions}
\label{sec:dm_inventory}


The total observed DM can be separated into contributions from three primary components,


\begin{equation}
\begin{split}
    \rm DM_{obs} = \dmism + \dmhalo + 
    DM_{EG}
    \\
    \rm DM_{EG} =  \dmcosmic  + \frac{DM_{host}}{1+z} \label{eq:dm}
\end{split}
\end{equation}

\noindent where \dmism\ is the contribution from the MW’s ISM and
\dmhalo\ is the contribution from the MW halo. $\mathrm{DM_{EG}}$ is the extragalactic DM contribution composed of \dmcosmic\ which is the contribution from the cosmic web (combined effects of the
intergalactic medium (IGM) and intervening galaxies), and
$\frac{\dmhost}{1+z}$ which is the redshifted contribution from the host galaxy's ISM including its halo and any gas in the immediate vicinity of the FRB source. We take $\dmism = 49.2$~\dmunits\ to be the average of the estimates from the NE2001 and YMW16 Galactic electron density models. For the MW halo contribution, we use \cite{PZ19} to estimate a value of $\sim 40$~\dmunits along the line-of-sight to \frbseven. For \dmcosmic, we use the Macquart relation to estimate the mean value given by,

%

\begin{eqnarray}
\dmacosmic  & = & \int\limits_0^z \frac{c \bar{n}_e(z^\prime) \, dz^\prime}{H_0 (1+z^\prime)^2 {\rm E}(z)} \label{eq:DMigm},\\
{\rm where} \; {\rm E}(z) & = & \sqrt{\Omega_M (1+z^\prime)^3 + \Omega_\Lambda}.
\label{eq:Ez}
\end{eqnarray}
$\bar n_e$ is the mean density of electrons given by,
\begin{eqnarray}
\bar{n}_e = \fdz\rho_{b}(z)m_{p}^{-1} \chi_{e} \label{eq:nebar}\\
= \fdz\rho_{b}(z)m_{p}^{-1} (1 - {\rm Y}_{\rm He}/2),
\end{eqnarray} 
\noindent in which $m_{p}$ is the mass of a proton, $\fdz$ is the fraction of cosmic baryons in diffuse ionized gas and $\rho_{b}$ is the mass density of baryons defined as,

\begin{eqnarray}
\rho_{b}(z) = \Omega_b\rho_{c,0}(1+z)^3
\end{eqnarray} 

\noindent where $\rho_{c,0}$ is the critical density and $\Omega_b$ is the baryon density parameter. $\chi_{e}= Y_{\rm{H}}+Y_{\rm{He}}/2 \approx 1-Y_{\rm{He}}/2$ is calculated from the primordial hydrogen and helium mass fractions $Y_{\rm{H}}$ and $Y_{\rm{He}}$. 

Using the algorithm encoded in the
FRB repository \citep{frbrepo}
and the Planck2018 cosmology \citep{planck2018},
we obtain $\dmcosmic = 121$~\dmunits. 
We combine the \dmism, \dmhalo, and \dmcosmic\, estimates with the total observed DM to constrain the 
host galaxy DM to be,  
$\frac{\dmhost}{1+z} = 361$.4~\dmunits. 

There is, however, considerable scatter
expected in \dmcosmic\ originating from
anisotropies in the cosmic web
\citep{mcquinn2014}.
Adopting the formalism introduced
by \cite{macquart+2020}, which describes
scatter in \dmcosmic\ with the 
$F$ parameter, and also their 
estimate of 0.32 for $F$, 
the 95\%\ interval on \dmcosmic\
is $\dmcosmic = [61, 346]$\,\dmunits.
This yields a 95\%\ c.l. range for 
$\frac{\dmhost}{1+z} = [140,425]$~\dmunits. 
This is fully consistent with the 
recent estimate for \dmhost\ from
\cite{james+2022} who modeled a 
population of FRBs including 16 with
redshift estimates.

\subsubsection{Constraining DM$_{\rm{host}}$ from $H\alpha$ measurements}
\label{sec:dminventory}


The DM contribution of the host galaxy may be independently estimated from its $\rm{H}_\alpha$ emission by converting the $\rm{H}_\alpha$ flux density,
$F_{\rm{H\alpha}} = 1.5\times 10^{-16}$ erg s$^{-1}$ cm$^{-2}$ to a ${\rm{H}}\alpha$ surface brightness of $S({\rm{H}}\alpha) = 2.74 \times 10^{-16}$ erg s$^{-1}$ cm$^{-2}$ arcsec$^{-2}$
assuming an angular size $\phi = 0.5\arcsec$. The $\rm{H}_\alpha$  surface brightness has been obtained by correcting the flux by Galactic extinction.
For a temperature $T = 10^4$ $T_4$ K, we express the emission measure, 
(\rm{EM} = $\int{n_{e}^{2}\, dl}$)
in the source frame,

\begin{equation}
\begin{split}
    {\rm{EM}}({\rm{H}}\alpha) = 4.858 \times 10^{17} \, T^{0.9}_{4} S({\rm{H}}\alpha)\, \rm{pc\,cm^{-6}} \\
    \approx 224.7\, \rm{pc\,cm^{-6}}.
\end{split}
\end{equation}

The host DM in the source frame derived from EM \citep{cws+16, tbc+17} is given by,

\begin{equation}
\begin{split}
    {\rm{DM}} = 387\,\dmunits {L^{1/2}_\mathrm{kpc}}
    \Bigg[\frac{4{\mathrm{f}}_{f}}{\upzeta(1+\epsilon^{2})}\Bigg] \times \Bigg(\frac{\mathrm{EM}}{600\,\dmunits}\Bigg) \\
    \approx 207.5 \, \dmunits,
\end{split}
\end{equation}

\noindent where $\upzeta \geqslant 1$ quantifies cloud-to-cloud variations in the ionized region of depth $L^{1/2}_\mathrm{kpc}$, with ${\rm{f}}_{f}$ being the volume filling factor of the ionized clouds. $\epsilon \leqslant 1$ is the fractional density variance inside discrete HII clouds \citep{cws+16, tbc+17}. Combining the host galaxy DM constraints of 361.4~\dmunits\, obtained after accounting for various other DM contributions in Section \ref{sec:dm_inventory}, and 207.5~\dmunits\, from $\rm{H}\alpha$ emission, we estimate that $\approx154$~\dmunits\, is unaccounted for. However, this excess is consistent within the 95\%\ c.l. range for 
$\frac{\dmhost}{1+z}$ in Section \ref{sec:dminventory} suggesting that it could well arise from the ISM of the host galaxy. 

\section{Discussion}
\label{sec:discussion}


\subsection{Non-detection of a PRS}

Our radio observations do not detect the presence of a PRS, suggestive of a nebula, at the position of \frbseven. Consequently, we place a 3$\sigma$ upper limit of $6.3\times 10^{27}$ erg s$^{-1}$ Hz$^{-1}$ on the luminosity at 1.284~GHz, which is nearly two orders of magnitude below the 1.77~GHz luminosity of the PRS associated
with FRB~20121102A \citep{mph+17}. If \frbseven\, is a repeater, it may be much older with an underluminous PRS compared to the more active repeaters like FRB~20121102A \citep{mph+17} and FRB~20190520B \citep{mnh+20} which are associated with compact PRSs, or it may be residing in a comparatively more diffuse environment like the repeaters FRB~20200120E \citep{kmn+22} and FRB~20201124A \citep{rll+22}, both of which lack a detectable PRS. 

\subsection{Host galaxy association and properties}

\begin{figure}
 \centering
\includegraphics[width=3.3in]{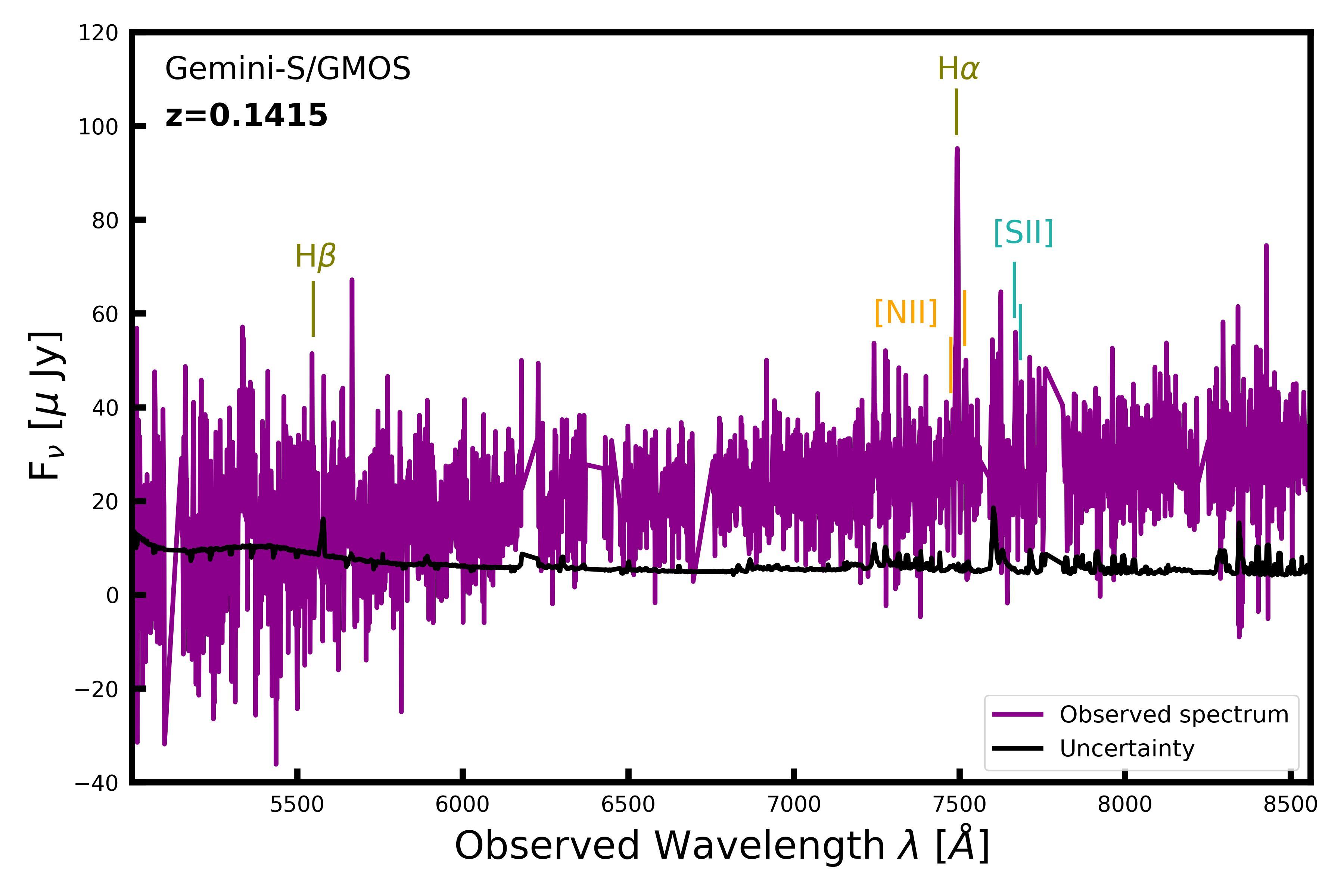}
\caption{\texttt{Prospector}-normalized spectrum \citep[see][for details on the normalization]{gfk+23} of \frbseven. Notable emission lines are denoted by colored lines: Balmer lines in green, nitrogen lines in yellow, and silicon lines in blue.}
\label{fig:spectrum} 
\end{figure}

We present the \texttt{Prospector}-normalized spectrum for the host of \frbseven\ in Figure~\ref{fig:spectrum} and in Table~\ref{tab:burstproperties} we summarize its key properties. The host galaxy is best described by a stellar mass of log(M$_*$/M$_{\odot}$) = $9.46^{+0.05}_{-0.06}$, slightly lower than that of the MW, but not low enough to be considered a dwarf galaxy. The present-day star formation rate of $0.03^{+0.03}_{-0.01}$ M$_{\odot}$ yr$^{-1}$ is low, indicating very little active star formation. Further details on the star formation history and specific prior ranges can be found in \cite{gfk+23}. 
We detect H$\beta$ emission, the [NII] doublet, and the [SII] doublet. There is a notable lack of [OIII] lines, which is a common feature in FRB host galaxies. Otherwise, the spectrum is typical of an FRB host galaxy \citep{gfk+23}. 


\begin{figure}
 \centering
\includegraphics[width=3.7in]{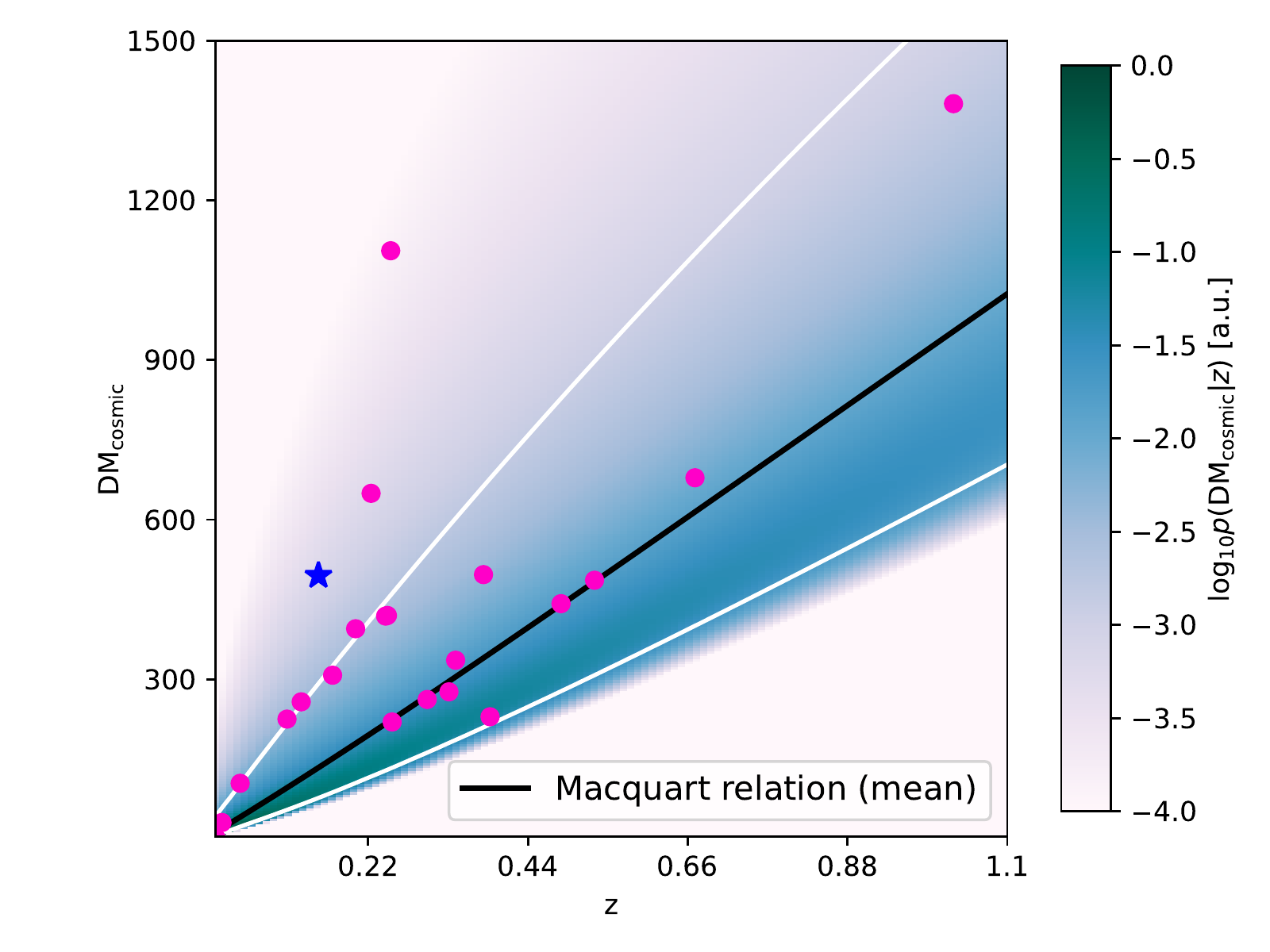}
\caption{The Macquart-relation for a sample of localised FRBs. DM$_\mathrm{cosmic}$ (after accounting for the MW ISM and 50 \dmunits\, for a MW halo contribution) is plotted as a function of the measured redshift for all current arcsecond- and subarcsecond-localised FRBs. The mean denotes the expected relation between DM$_\mathrm{cosmic}$ and redshift for a universe based on the Planck cosmology. The white solid lines encompass 90\% of the DM$_\mathrm{cosmic}$ values from a model for ejective feedback in Galactic haloes that is motivated by simulations with feedback parameter $F = 0.32$. FRBs with significant host and/or local DM contributions are seen to deviate significantly from the scatter around the relation. \frbseven\, with the MW ISM and MW halo DM contributions accounted for, is denoted by the star.}
\label{fig:macquart-relation} 
\end{figure}

\input{tab_photo-z}
\begin{figure*}
    \centering
    \includegraphics[width=\textwidth]{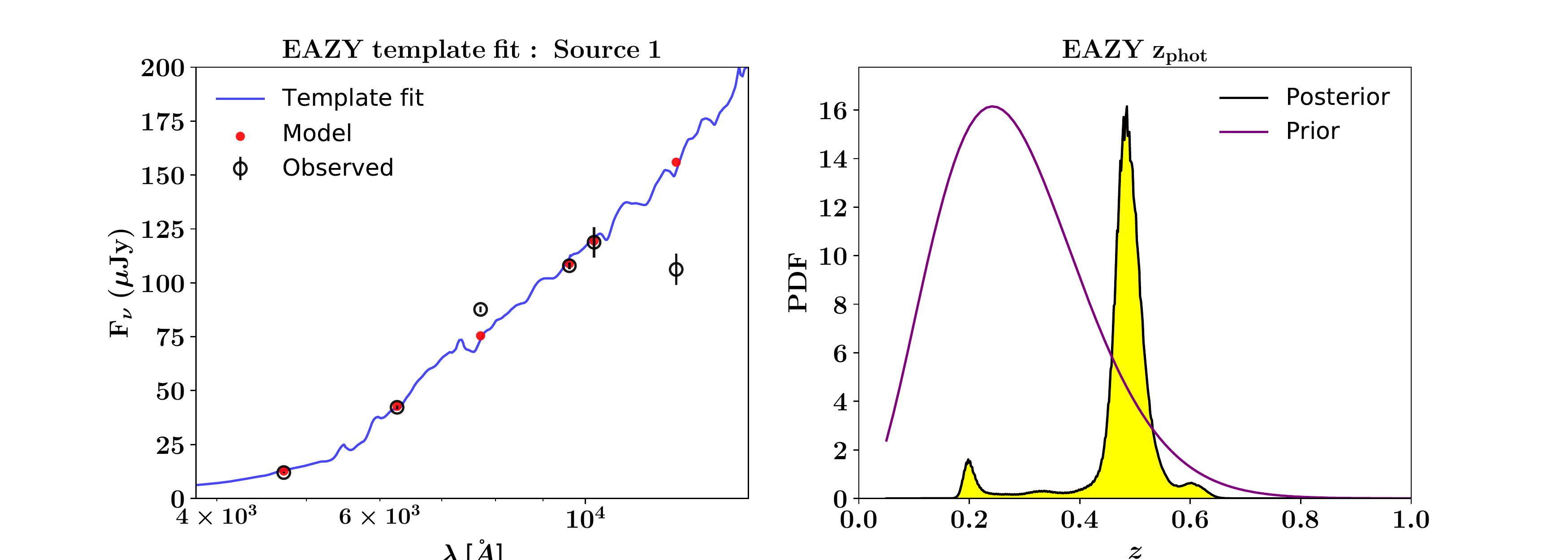}
    \caption{An example photometric redshift estimation for Source~1 using EAZY. \emph{Left}: Best-fit template spectrum (blue) with input photometry (black) overlaid. The red points show synthetic photometry from the best-fit template. \emph{Right}: The resultant $z_\mathrm{phot}$ probability distribution function (PDF). The broad, multimodal posterior distribution implies more data points are necessary to better constrain the galaxy spectrum better. This is the case for all sources analyzed (see Table \ref{tab:photoz} for a summary).}
    \label{fig:eazy}
\end{figure*}

\begin{figure}
 \centering
\includegraphics[width=3.8in]{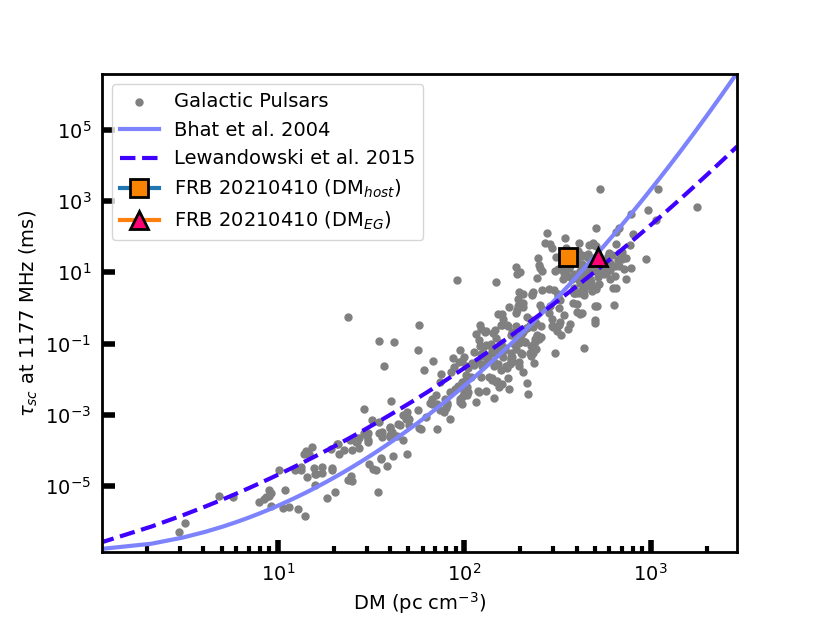}
\caption{Dispersion measure as a function of pulse scattering timescale for pulsars in the Galaxy. The extragalactic and the host DM contributions to~\frbseven~ are marked with a triangle and a square. The solid line corresponds to the best-fit relation between scattering and DM from the Galactic pulsars as presented in \citet{bcc+04} while the dashed line is the best fit from~\citet{lkk15}}.
\label{fig:DM_scatter} 
\end{figure}

\subsection{Host galaxy DM contribution}

We consider different possible scenarios for the origin of the host galaxy DM. It could arise from, 1) gas in the circumburst region of the progenitor, 2) the ISM of the host galaxy or 3) gas in an intervening foreground halo. The absence of a PRS and strong Faraday rotation in \frbseven\, differs from other FRBs which exhibit significant host DMs \citep{nal+22}, indicating that it is not associated with a complex and highly dynamic magneto-ionic plasma. This is the first FRB that is not associated with a dwarf galaxy to exhibit a significant host galaxy DM contribution \citep{tbc+17, nal+22, bgs+22}, and seen to deviate beyond the 90\% c.l. of the Macquart relation~\citep{macquart+2020} as shown in Figure \ref{fig:macquart-relation}. It is likely that the host galaxy DM contribution we observe for \frbseven\, arises from the ISM of the host galaxy rather than the immediate environment or circum-burst medium of the progenitor \citep{yly+22}. This is consistent with the observed large scattering timescale, which is expected to arise from a scattering screen located in the host galaxy. As the morphology of the galaxy remains unclear, it is impossible to determine if it is an edge-on system. In any case, it is plausible that the FRB sightline passed through high-density clumps of gas, like HII regions in the ISM of the host which could contribute to the host galaxy DM. 

\subsection{Photometric redshifts}
We estimated photometric redshifts, $z_\mathrm{phot}$, for the sources in Figure~\ref{fig:optical_galaxies} using the EAZY template fitting code \citep{eazy}. We fit flux measurements in the SOAR \emph{griz} and VISTA \emph{YJ} bands with the \texttt{CWW+KIN} set while allowing superpositions of the individual templates. Although VISTA \emph{Ks} band data was available, we excluded them from the fits as those fluxes appeared to be consistently $\sim2$ times higher than the expected value from the template fits. The photometric redshifts thus obtained are listed in Table \ref{tab:photoz}. An example fit is shown in Figure \ref{fig:eazy}. We note that the host $z_\mathrm{phot}$ is not entirely consistent with the secure spectroscopic redshift. In general, this, along with the wide photometric redshift error margins for all sources in the field imply the need for flux measurements in more bands if not spectroscopic observations of all sources for more accurate redshifts. 


\subsection{Pulse morphology}

It can be seen in Figure \ref{fig:scattering}, that our pulse profile model seems to require a shifted low frequency emission region, which suggests a subpulse scenario. This time-frequency downward drifting structure seen in \frbseven\, is suggestive of a repeater origin when compared against the large sample of FRBs in \cite{pgk+21}.

In a pulse profile, the effect of thin-screen scattering appears as a truncated exponential or `exponential tail', which is a broadening of the trailing edge (or `tail') of the profile \citep{LT1975}. The scattering measured for this FRB shows that it is dominated by the scattering contribution from the host galaxy (see Section \ref{sec:scattering}). This is further highlighted if we study the correlation between the host contribution to the DM and the total amount of scattering. Figure~\ref{fig:DM_scatter} compares the scattering time as a function of DM for pulsars in the MW with that of \frbseven. It is clear that for the host contribution to the total DM, the scattering timescale is consistent with what one would obtain for an object in our Galaxy~\citep{bcc+04,lkk15}. This shows that the turbulent environment in the host galaxy is similar to that observed in the plane of our own Galaxy. 


\section{Summary}
\label{sec:conclusion}

We present the discovery and sub-arcsec localisation of \frbseven\, with the MeerKAT radio telescope. \frbseven\, is the first one to exhibit a DM$_\mathrm{cosmic}$ which deviates from the Macquart relation while not being associated with a dwarf galaxy. The host galaxy mass is only slightly lower than that of the MW. While the potential downward frequency drifting in the spectrum is reminiscent of repeaters \citep{pgk+21}, we are yet to detect a repeat pulse from this source. Based on the absence of a PRS we place a 3$\sigma$ upper limit of $6.3\times 10^{27}$ erg s$^{-1}$ Hz$^{-1}$ on the luminosity at 1.284~GHz. We investigated the origin of the observed scattering in \frbseven\, and find that it cannot originate from a screen in the MW or the halo of an intervening galaxy. Additionally, the lack of a PRS associated with the FRB means that the scattering is most likely dominated by turbulent material in the ISM of the host galaxy. The absence of strong Faraday rotation in \frbseven\, sets it apart from other FRBs with large values of DM$_\mathrm{cosmic}$. We encourage follow-up observations to search for repeating pulses.

\section*{Acknowledgements}
M.C. would like to thank Clancy W. James for help with the {\sc zdm} code. The authors would also like to thank the South African Radio Astronomy Observatory staff for their help with scheduling the observations. M.C., B.W.S., K.M.R., T.B., L.N.D., S.S., M.M., V.M., M.S. and F.J. acknowledge funding from the European Research Council (ERC) under the European Union's Horizon 2020 research and innovation programme (grant agreement No. 694745). M.C. acknowledges support of an Australian Research Council Discovery Early Career Research Award (project number DE220100819) funded by the Australian Government and the Australian Research Council Centre of Excellence for All Sky Astrophysics in 3 Dimensions (ASTRO 3D), through project number CE170100013. K.M.R. acknowledges support from the Vici research program `ARGO' with project number 639.043.815, financed by the Dutch Research Council (NWO).
Authors W.F., C.K., J.X.P., A.C.G., S.S. and N.T. as members of the Fast and Fortunate for FRB Follow-up team, acknowledge support from NSF grants AST-1911140 and AST-1910471. N.T. and L.B. acknowledge support by FONDECYT grant 11191217. 
The MeerKAT telescope is operated by the South African Radio Astronomy Observatory (SARAO), which is a facility of the National Research Foundation, an agency of the Department of Science and Innovation. The Parkes Radio Telescope (Murriyang) is managed by CSIRO. The FBFUSE beamforming cluster was funded, installed and operated by the
Max-Planck-Institut f\"{u}r Radioastronomie and the Max-Planck-Gesellschaft. We acknowledge the Wiradjuri people as the traditional owners of the Parkes observatory site. We acknowledge the use of the ilifu cloud computing facility - www.ilifu.ac.za, a partnership between the University of Cape Town, the University of the Western Cape, the University of Stellenbosch, Sol Plaatje University, the Cape Peninsula University of Technology and the South African Radio Astronomy Observatory. The ilifu facility is supported by contributions from the Inter-University Institute for Data Intensive Astronomy (IDIA - a partnership between the University of Cape Town, the University of Pretoria and the University of the Western Cape), the Computational Biology division at UCT and the Data Intensive Research Initiative of South Africa (DIRISA).

Based on observations obtained at the international Gemini Observatory, a programme of NSF’s NOIRLab, which is managed by the Association of Universities for Research in Astronomy (AURA) under a cooperative agreement with the National Science Foundation on behalf of the Gemini Observatory partnership: the National Science Foundation (United States), National Research Council (Canada), Agencia Nacional de Investigaci\'{o}n y Desarrollo (Chile), Ministerio de Ciencia, Tecnolog\'{i}a e Innovaci\'{o}n (Argentina), Minist\'{e}rio da Ci\^{e}ncia, Tecnologia, Inova\c{c}\~{o}es e Comunica\c{c}\~{o}es (Brazil), and Korea Astronomy and Space Science Institute (Republic of Korea). The Gemini data were obtained from programs GS-2021B-Q-138 and GS-2022A-Q-143 (PI Tejos), and were processed with Pypeit \citep{pypeit}.

Based in part on observations obtained at the Southern Astrophysical Research (SOAR) telescope, which is a joint project of the Minist\'{e}rio da Ci\^{e}ncia, Tecnologia e Inova\c{c}\~{o}es (MCTI/LNA) do Brasil, the US National Science Foundation’s NOIRLab, the University of North Carolina at Chapel Hill (UNC), and Michigan State University (MSU).

Based on observations obtained as part of the VISTA Hemisphere Survey, ESO Progam, 179.A-2010 (PI: McMahon)

A portion of this research was performed at the Jet Propulsion Laboratory, California 
Institute of Technology, under a Research and Technology 
Development Grant through a contract with the National Aeronautics and Space 
Administration. U.S. government sponsorship is acknowledged.

\section*{Data Availability}
The data will be made available upon reasonable request to the authors.



\bibliographystyle{mnras}
\bibliography{bib.bib} 








\bsp	
\label{lastpage}
\end{document}

%% file: tab_path.tex
\begin{table*}
\caption{FRB PATH Associations\label{tab:path}}
\begin{tabular}{cccccccc}
Name & RA$_{\rm cand}$ & Dec$_{\rm cand}$ 
& $\theta$ 
& \halflight  
& mag 
& \PO & \POx 
\\& (deg) & (deg) & ($''$) & ($''$)  
\\ 
\hline 
\hline 
\hostname\ & 326.0862 & $-79.3180$& 0.78& 0.83& 21.78& 0.1611& 0.9957\\ 
Source 1 & 326.0793 & $-79.3180$& 4.69& 0.94& 20.36& 0.6315& 0.0043\\ 
Source 2 & 326.0948 & $-79.3184$& 5.74& 0.50& 22.79& 0.0657& 0.0000\\ 
Source 3 & 326.0786 & $-79.3187$& 5.39& 0.68& 21.92& 0.1417& 0.0000\\ 
\hline 
\end{tabular} 
\end{table*}

%% file: tab_photo-z.tex
\begin{table}
\centering
\caption{EAZY photometric redshifts of sources\label{tab:photoz}}
\begin{tabular}{cccc}
Name & $z_\mathrm{phot}$ & 68\% c.i. & 95\% c.i.\\ 
\hline 
\hline 
\hostname\ & 0.44 & 0.41-0.50 & 0.22-0.54\\ 
Source 1 & 0.47 & 0.45-0.52 & 0.20-0.59\\ 
Source 2 & 0.58 & 0.53-0.64 & 0.50-0.65\\ 
Source 3 & 0.53 & 0.44-0.63 & 0.21-0.68\\ 
\hline 
\end{tabular} 
\end{table} 